\documentclass[a4paper,10pt,twoside]{cpc-hepnp}
\usepackage{CJKutf8}
\pdfoutput=1 % if your are submitting a pdflatex (i.e. if you have
\usepackage{CJK,upgreek,fancyhdr}
\usepackage{multicol}
\usepackage{graphicx, subfigure}
\usepackage{booktabs}
\usepackage{amssymb,bm,mathrsfs,bbm,amscd}
\usepackage[tbtags]{amsmath}
\usepackage{lastpage}
\usepackage[usenames, dvipsnames]{color}
\usepackage{soul}

\begin{document}
%\begin{CJK*}{GB}{gbsn}an   
%\begin{CJK*}{GBK}{song}

\fancyhead[co]{\footnotesize Development of an ADC Radiation Tolerance Characterization System for the Upgrade of the ATLAS LAr Calorimeter}
\fancyfoot[C]{\small 010201-\thepage}

%\footnotetext[0]{Received 31 June 2015}

\title{Development of an ADC Radiation Tolerance Characterization System for the Upgrade of the ATLAS LAr Calorimeter
\thanks{Supported by the U.S. Department of Energy under Contract No. DE-SC001270.}}

\author{%
	Hong-Bin Liu
	\begin{CJK}{UTF8}{gbsn}(刘洪斌)\end{CJK}$^{1,2,3,1)}$\email{hliu2@bnl.gov}
	\quad Hu-Cheng Chen
	\begin{CJK}{UTF8}{gbsn}(陈虎成)\end{CJK}$^{3}$
	\quad Kai Chen
	\begin{CJK}{UTF8}{gbsn}(陈凯)\end{CJK}$^{3}$
	\quad James Kierstead$^{3}$\\
	\quad Francesco Lanni$^{3}$
	\quad Helio Takai $^{3}$
	\quad Ge Jin
	\begin{CJK}{UTF8}{gbsn}(金革)\end{CJK}$^{1,2}$
}
\maketitle

\address{
	$^1$State Key Laboratory of Particle Detection and Electronics, University of Science and Technology of China, Hefei Anhui 230026, China\\
	$^2$Department of Modern Physics, University of Science and Technology of China, Hefei Anhui 230026, China\\
	$^3$Brookhaven National Laboratory, Upton NY 11973, U.S.A.\\
}

\begin{abstract}
ATLAS LAr calorimeter will undergo its Phase-I upgrade during the long shutdown (LS2) in 2018, and a new LAr Trigger Digitizer Board (LTDB) will be designed and installed. Several commercial-off-the-shelf (COTS) multi-channel high-speed ADCs have been selected as possible backups of the radiation tolerant ADC ASICs for the LTDB. To evaluate the radiation tolerance of these backup commercial ADCs, we developed an ADC radiation tolerance characterization system, which includes the ADC boards, data acquisition (DAQ) board, signal generator, external power supplies and a host computer. The ADC board is custom designed for different ADCs, with ADC drivers and clock distribution circuits integrated on board. The Xilinx ZC706 FPGA development board is used as a DAQ board. The data from the ADC are routed to the FPGA through the FMC (FPGA Mezzanine Card) connector, de-serialized and monitored by the FPGA, and then transmitted to the host computer through the Gigabit Ethernet. A software program has been developed with Python, and all the commands are sent to the DAQ board through Gigabit Ethernet by this program. Two ADC boards have been designed for the ADC, ADS52J90 from Texas Instruments and AD9249 from Analog Devices respectively.  TID tests for both ADCs have been performed at BNL, and an SEE test for the ADS52J90 has been performed at Massachusetts General Hospital (MGH). Test results have been analyzed and presented. The test results demonstrate that this test system is very versatile, and works well for the radiation tolerance characterization of commercial multi-channel high-speed ADCs for the upgrade of the ATLAS LAr calorimeter. It is applicable to other collider physics experiments where radiation tolerance is required as well.
\end{abstract}


\begin{pacs}
81.40.Wx, 07.05.Hd 
%1---3 PACS codes (Physics and Astronomy Classification Scheme, http://www.aip.org/pacs/pacs.html/)
\end{pacs}

%\footnotetext[0]{\hspace*{-3mm}\raisebox{0.3ex}{$\scriptstyle\copyright$}2013
%Chinese Physical Society and the Institute of High Energy Physics
%of the Chinese Academy of Sciences and the Institute
%of Modern Physics of the Chinese Academy of Sciences and IOP Publishing Ltd}%

\begin{multicols}{2}

\section{Introduction}
\label{sec:intro}
In order to support and extend the physics programs and discovery reach of the ATLAS experiment at LHC, ATLAS will have a long shutdown and Phase-I upgrade in 2018. A new LAr Trigger Digitizer Board (LTDB) will be installed in the available spare slots of the 
front-end 
crates of the Liquid Argon (LAr) calorimeter. The analog to digital converter (ADC) is a key electronic component of LTDB since there are 320 channels of analog signals from front-end, and all of them need to be digitized on the LTDB \cite{1}. Due to the critical radiation environment inside the LAr calorimeter, long-term exposure to ionizing radiation will cause parametric degradation and ultimately functional failure in electronic devices, so all the electronic devices installed inside the LAr calorimeter should be radiation tolerant. To fulfill the radiation tolerance requirements of the LTDB, radiation tolerant ADC ASIC designs are being developed. In the meantime, several commercial ADCs have been chosen as possible backups for the ADC ASIC.\par 
Commercial ADCs are not specifically designed for a radiation environment, so their radiation tolerance specifications are not available. In order to assess whether these commercial ADC candidates are suitable for the LTDB design of the LAr calorimeter upgrade, a generic, flexible radiation tolerance characterization system for multi-channel high-speed ADCs is developed.\par

In this paper, the test methods to perform the total ionizing dose (TID) test and single event effect (SEE) test are first described in Section 2. Section 3 provides an overview of the whole radiation characterization system, followed by a description of the hardware design of the ADC boards, Xilinx ZC706 \cite{2} development board based FPGA firmware design, and host software design in Sections 4 to 6. Finally, the test setup and results of the TID and SEE tests of two commercial ADCs are presented in Sections 7 and 8.

\section{Test Methods}
\label{sec:test_method}
\subsection{TID Test}
TID is a long-term effect of radiation on an electronic device due to the cumulative energy deposited in the material. Typical effects of TID include  parametric degradation, or variations in device parameters such as leakage current, threshold voltage or functional failures. Specific to ADC devices, the signal-to-noise ratio (SNR) and power consumption are two essential and testable parameters.\par
For an ADC device, the radiation tolerance is characterized by tracking two key parameters, the power consumption and signal-to-noise ratio (SNR) of the digitized data, according to the accumulation of the total ionization dose.\par
External linear power supplies are used to provide the necessary stable power to the ADCs under test. The current consumption of every power rail of the power supplies is recorded to study the trend of power consumption versus total dose. In the meantime, a sine-wave signal is generated by the signal generator and fed into the ADC for digitization. The characterization system will capture 16 thousand sampling points of digitized data every minute during the TID test, and store them in the host computer. An offline FFT calculation is used to check the SNR degradation tendency as the total dose accumulates.

\subsection{SEE Test}
A SEE in an ADC is a change of state caused by the strike of a single ionizing particle, which can deposit sufficient energy to the sensitive nodes in the ADC.  There are many types of SEE and they can be divided into two main categories: soft error and hard error \cite{3}.\par 

In this paper, the SEEs of the ADC are divided into three categories: single event upset (SEU), single event function interrupt type A (SEFI-A) and single event function interrupt type B (SEFI-B). SEU refers to bit-flip events and can be categorized as a soft error. SEFI refers to a function interrupt of the device, from which the device cannot recover to its correct working state automatically. Type A SEFIs are function interrupts that can be recovered by a hardware reset, and type B SEFI are function interrupts that need a power recycle to bring the device back to its normal operational status.
\par 

As shown in the SEE test flow chart of Figure~\ref{fig:see_flow_chart}, the first step after the start of the SEE test is the LUT calculation and configuration. The host computer will capture 16 thousand points of output data from all the ADC channels, then calculate the average waveforms and configure them into a lookup table (LUT). After the phase adjustment and threshold setting, the FPGA continuously checks the difference between the output of the LUT and ADC. If the difference exceeds the pre-set threshold, an SEE event will be flagged and the test beam will been stopped. Sixteen thousand points of ADC data from all the ADC channels are recorded, including the data of 1024 sampling points before the point where the SEE event is flagged.\par
\begin{center}
		\centering
		\includegraphics[width={0.24\textwidth}]{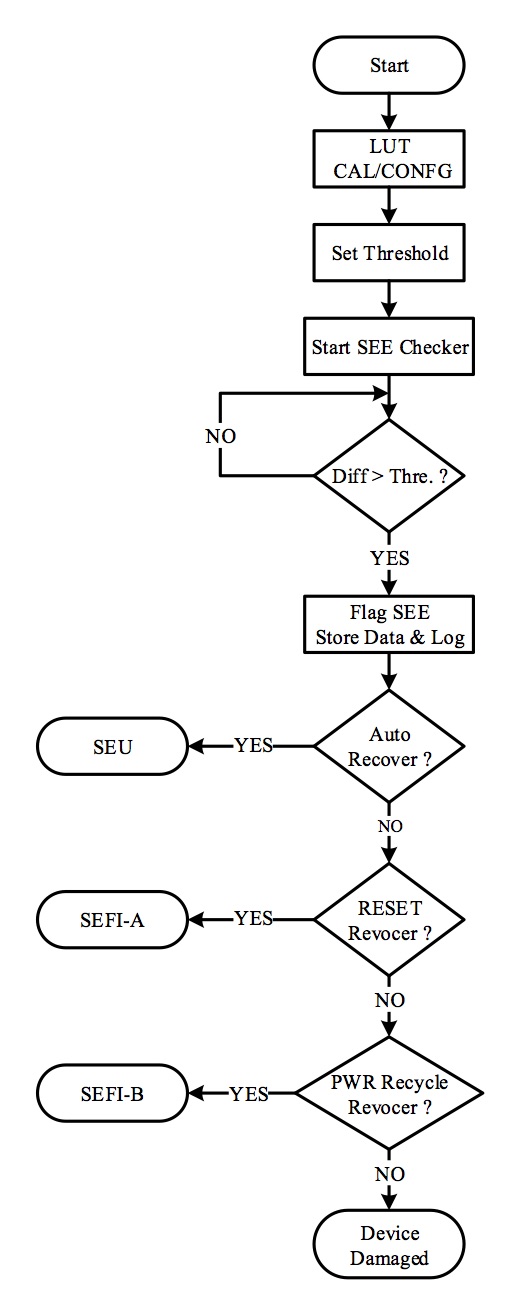}
		\figcaption{\label{fig:see_flow_chart}Flow chart of the SEE test method.}
\end{center}
The event type of the flagged SEE event is determined by the approaches that can bring the device back to its normal status as described before. 
\par 
After the flagged SEE event is categorized and the device is back to its normal operational status, the SEE checker starts again after the LUT has been re-calculated and re-configured.

\section{System Overview}
\label{sec:sys_overview}
Figure~\ref{fig:sys_bd} shows the block diagram of the ADC radiation tolerance characterization system. The system comprises several power supplies with the Ethernet GPIB controller, a signal generator with  splitter, the DUT (Device Under Test, the ADC) board, DAQ board, an Ethernet-controlled power outlet, an Ethernet router, and a host computer.\par
\end{multicols}

\begin{center}
		\centering
		\includegraphics[width=12cm]{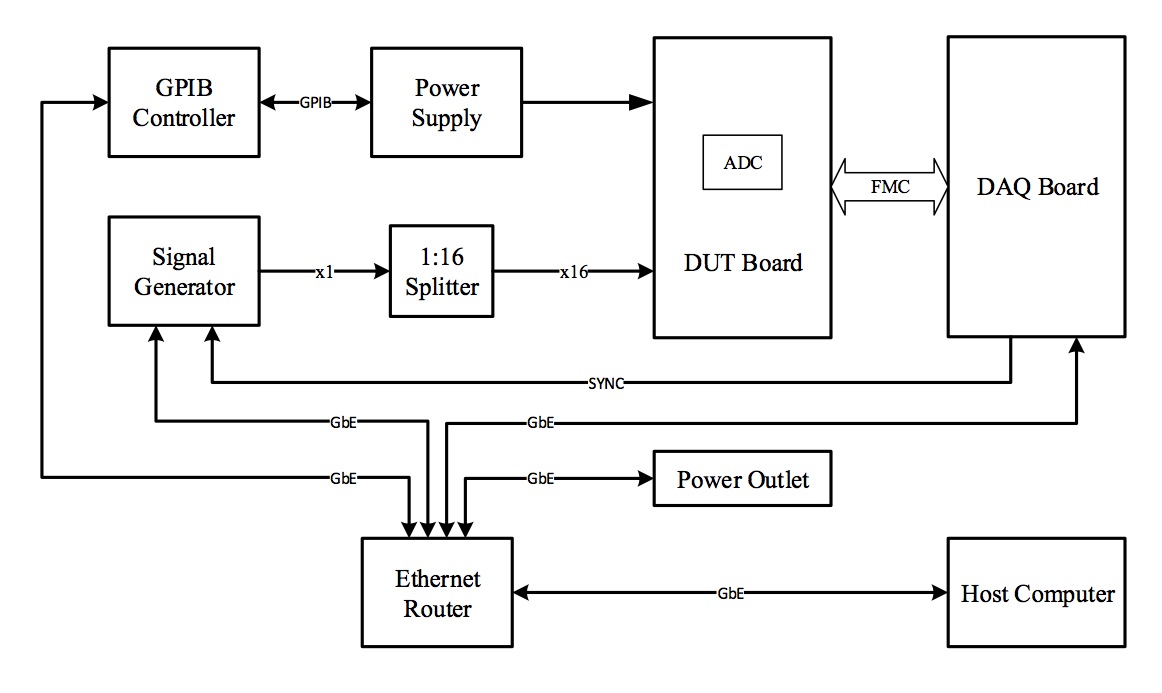}
		\figcaption{\label{fig:sys_bd}Block diagram of the characterization system.}
\end{center}

\begin{multicols}{2}
The analog signal is generated by the signal generator and then split into 16 analog signals, which are fed into the ADC under test. To make the signal synchronized to the FPGA clock, a 10 MHz SYNC clock is generated by the FPGA and sent to the signal generator.\par 
In this system, external power supplies are used to provide multiple power rails to the ADC under test. Since the external power supplies can be shielded separately, the error caused by the degradation of the power supply can be eliminated compared to the onboard voltage regulator solution. The current and voltage of all the power rails are monitored and recorded every minute by the host computer through the GPIB interface during the test period.\par 
The DUT board is an ADC-specific custom designed FMC mezzanine board. The ADC and its driver circuits are implemented on this board. Analog signals from the splitter are fed into this board, then all of them are digitized by the ADC. The digital outputs of the ADC are routed to the FMC connector directly. An FMC extension cable is used to connect the ADC board to the DAQ board.\par
The DAQ board is the core of this ADC characterization system, and it is based on the Xilinx ZC706 FPGA development board. All the digital data from the ADC are received and processed by the DAQ board. The associated TID and SEE test methods mentioned in Section 2 are implemented on this board.\par 
The host computer is used to run the test software and issue all the control commands. All essential data and logs from the DAQ board are transferred to the host computer and stored on its hard drive for offline analysis. The Ethernet router is used to connect all components with Ethernet interface and the host computer.\par
In addition, the power for all the components of this system are connected to a power outlet which can be controlled remotely through the Ethernet interface. The power recycles of all the components in this system can be easily performed by configuring the power outlet.

\section{ADC Board Design}
\label{sec:adc_brd_design}
\begin{figure} 
\centering 
\subfigure[~]{\includegraphics[width={0.45\textwidth}]{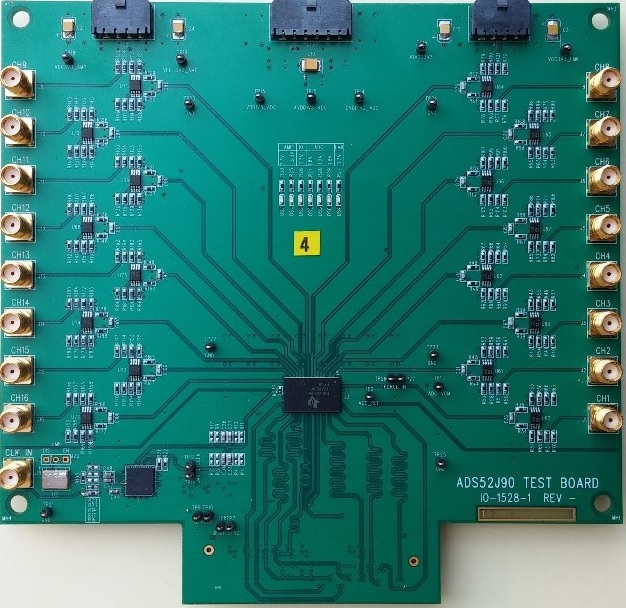}\label{fig:ads52j90_brd} }
\subfigure[~]{\includegraphics[width={0.45\textwidth}]{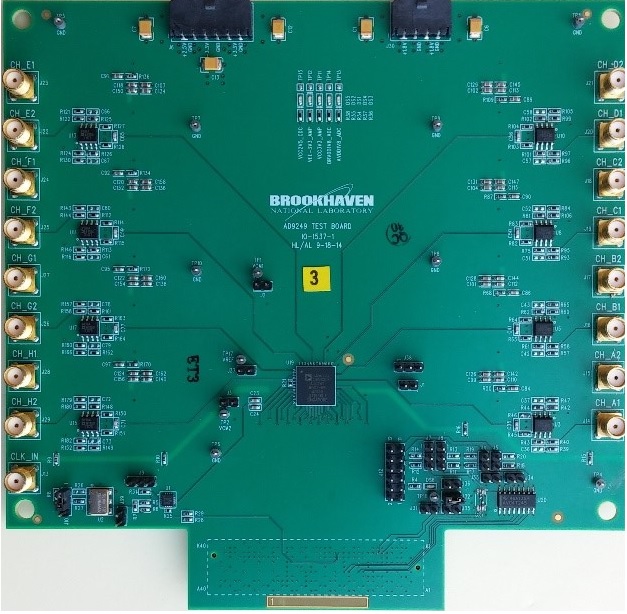} \label{fig:ad9249_brd} } 
\caption{Photos of the ADC board: (a) ADC board for the ADS52J90 from Texas Instruments; (b) ADC board for the AD9249 from Analog Devices.} 
\label{fig:adc_board_photo} 
\end{figure}

\begin{figure} 
\centering 
\subfigure[~]{\includegraphics[width={0.5\textwidth}]{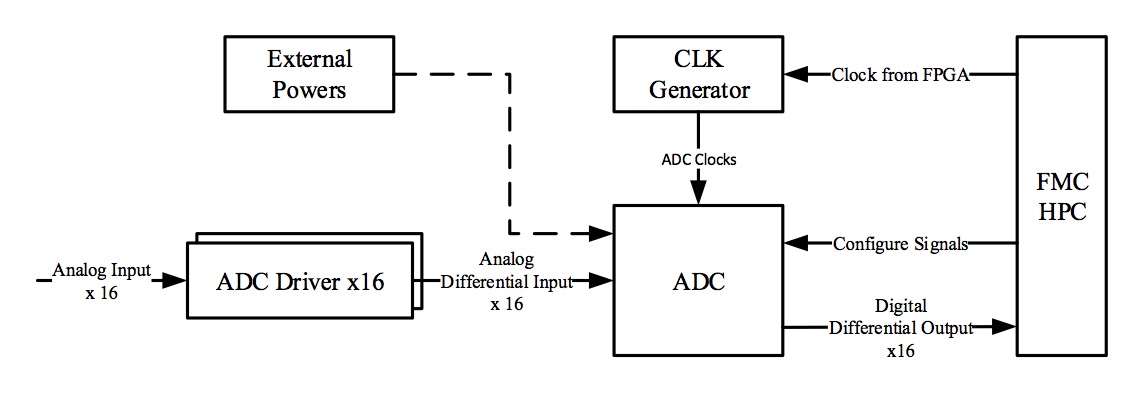}\label{fig:adc_boards_bd} }
\subfigure[~]{\includegraphics[width={0.5\textwidth}]{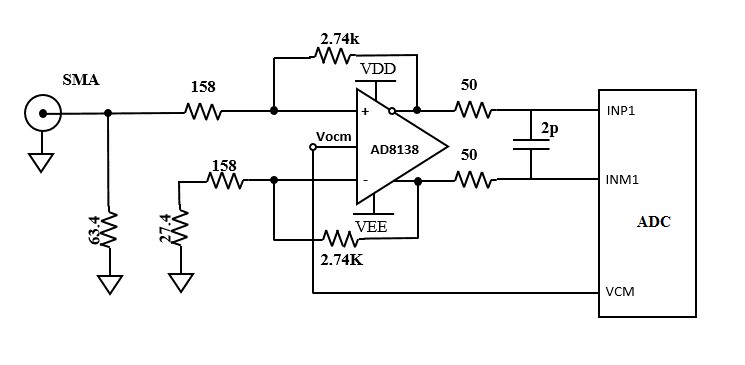} \label{fig:adc_driver_sch} } 
\caption{(a) Block diagram of the ADC boards. (b) Simplified schematic of the ADC driver circuit.} 
\label{fig:adc_board_design} 
\end{figure}

Two similar ADC boards have been designed, for the ADS52J90 \cite{4} from Texas Instruments and AD9249 \cite{5} from Analog Devices respectively. Photos of these two boards are shown in Figure~\ref{fig:adc_board_photo}. Both ADC boards have similar architecture, which is shown in Figure~\ref{fig:adc_boards_bd} as a simplified block diagram. The ADC boards comprise 16 channels of ADC driver circuit, a clock generator/buffer chip, the ADC, and external power connectors.\par

The clock requirements of the ADS52J90 and AD9249 are different, so different clock schemes are implemented. For the ADS52J90, it has a JESD204B interface, which requires a frame clock and a sampling clock from the same clock source with different frequency. The clock generator LMK03200 from TI is implemented to produce these clocks from a source clock generated by the FPGA. For the AD9249, it only needs a sampling clock, so a 2:4 clock buffer is used to select the clock source for the ADC.\par

All the digital outputs and control signals of the ADC are connected to the DAQ board through the FMC connector. The FMC connector is a 400-pin high-density connector, and the pin-out of all the signals is compatible with the VITA 57.1 standard \cite{6}.\par

The circuit of the ADC driver is shown in Figure~\ref{fig:adc_driver_sch}. It is a single-end to differential amplifier. The output of the differential amplifier is DC coupled to the ADC. The common-mode voltage of the differential output is determined by the output-common voltage of the ADC.

\end{multicols}

\begin{multicols}{2}

\section{FPGA Firmware}
\label{sec:fw}
The block diagram of the FPGA firmware is shown in Figure~\ref{fig:firmware_bd}. The firmware is compatible with both ADC boards, and two different kinds of ADC interface are implemented.\par

\end{multicols}

\begin{center}
		\centering
		\includegraphics[width={0.8\textwidth}]{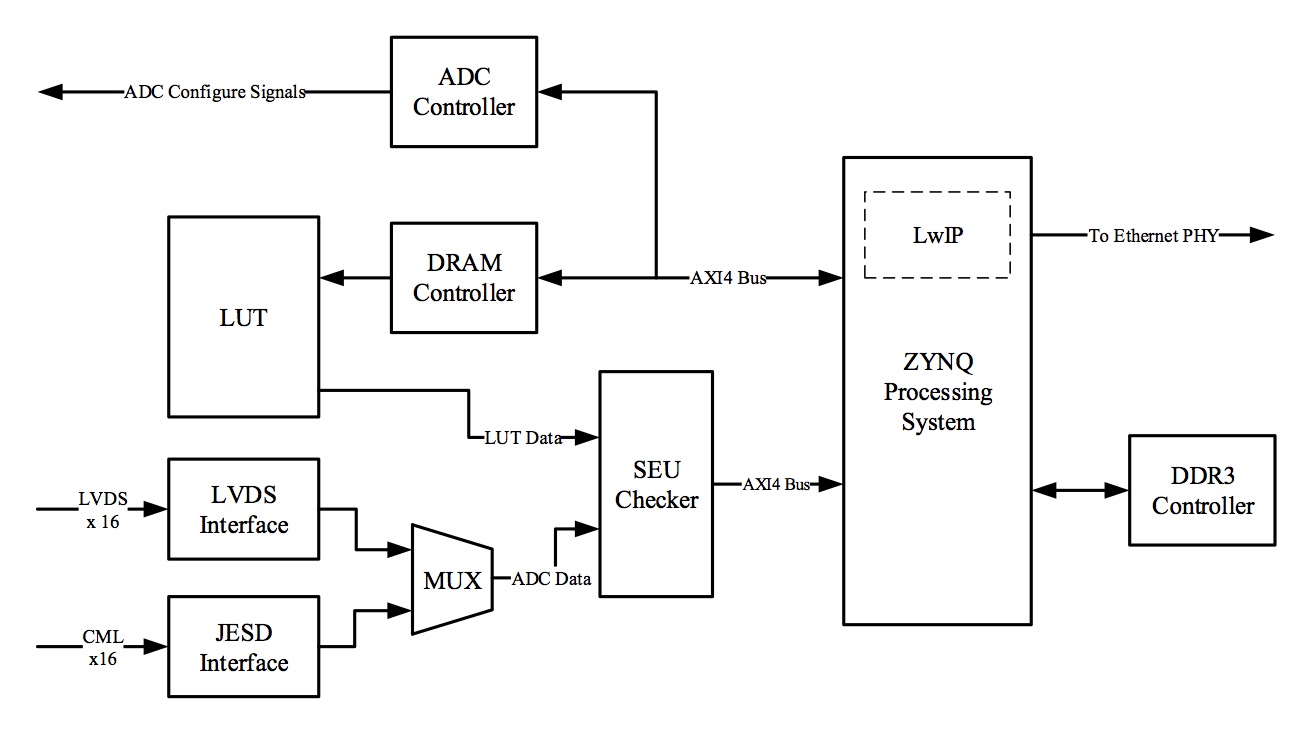}
		\figcaption{\label{fig:firmware_bd} Block diagram of the ZYNQ FPGA firmware.}
\end{center}

\begin{multicols}{2}

The LUT is built with internal block memory, and the contents of the LUT are calculated by the host computer according to the data acquired from the ADC.\par
When the test is started, the data from the LUT is synchronized with the ADC data and sent to the SEU checker. The SEU checker will check the difference of the data from ADC and LUT, and an SEE event will be flagged when the difference is bigger than the programmable threshold.
\par 
The SEE event flag will trigger a data readout procedure, where 16 thousand points of the ADC data and the LUT data are buffered into the DDR3 memory and read out by the host computer. All the data are stored on the hard drive for offline analysis.
\par
In addition, an ADC controller is implemented to configure the ADS52J90 and AD9249 through the SPI and uWIRE interface respectively.\par
A light-weight IP (LWIP) protocol \cite{7} is implemented in the processing system of the ZYNQ FPGA on the DAQ board ZC706. The host computer can talk to the system through the Ethernet link. The ZYNQ processing system will decode the commands from the host computer and route them to different function modules through the AXI4 bus.

\section{Host Software}
\label{sec:sw}
The host software is responsible for sending commands to the DAQ board and reading all the essential data from the DAQ board. The software is developed with Python and its GUI is based on the PYQT library. The host software works as a TCP/IP client of the DAQ board, and communicates with the DAQ board through TCP/IP sockets.

\section{Test Setup}
\label{sec:test_setup}
The TID test was performed by using the $^{60}Co$ source at the Solid State Gama-Ray Irradiation Facility at BNL. The dose rate at the position where the ADC boards were mounted is about 10 kRad(Si)/hour. The accurate dose rate was determined by optically stimulated luminescence (OSL) dosimeters provided and read by Landauer Inc. with an error of $\pm$6\%.\par 
The high-energy proton facility used to perform the SEE test is the proton beam facility of Francis H. Burr Proton Therapy Center at Massachusetts General Hospital (MGH) \cite{8} located in Boston. This is a proton beam facility with an energy that can be adjusted from 70 MeV to 230 MeV for a circular beam spot size of approximately 2.5 cm in diameter, and a flux that can be tuned from $5 \times 10^{7} $ to $1 \times 10^{9} p \cdot cm^{-2} \cdot s^{-1}$ with a variance in flux of approximately 10\% across the spot \cite{8}. The flux was fixed at 5 $\times 10^{7} p \cdot cm^{-2} \cdot s^{-1}$ during the SEE test.
\par
The components of the system setup are similar for the TID test and SEE test, except for the shielding method. Many lead bricks were used to shield our DAQ board, power supplies, signal generators, and router during the TID test, while polyethylene bricks were used to do the shielding when performing the SEE test.\par
\begin{figure} 
\centering 
\subfigure[~]{\includegraphics[width={0.4\textwidth}]{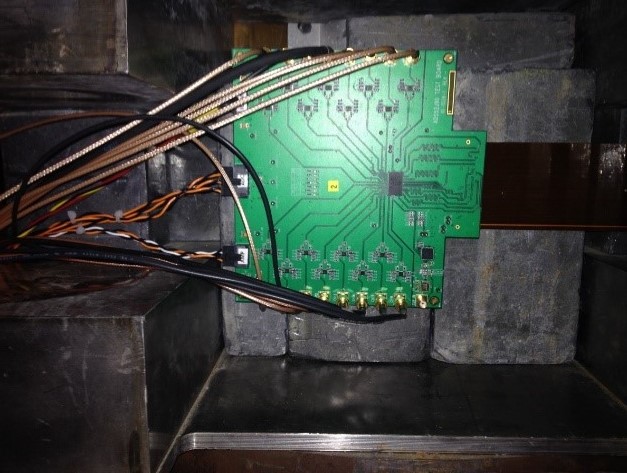}\label{fig:tid_field} }
\subfigure[~]{\includegraphics[width={0.4\textwidth}]{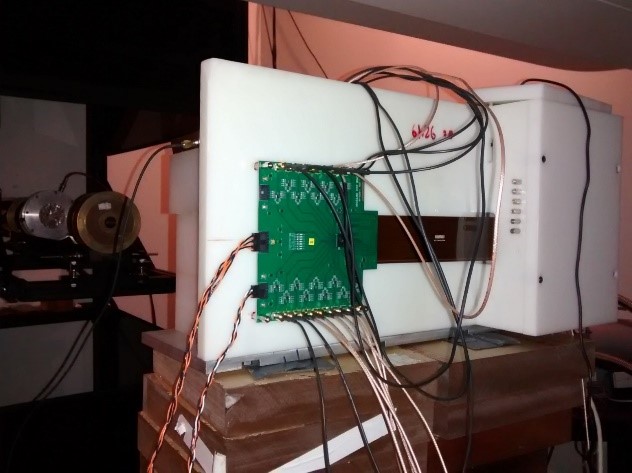} \label{fig:see_field} } 
\caption{Field photos of the TID test at BNL and SEE test at MGH. (a) Lead bricks were used to shield all the components except the ADC under test during the TID test. (b) Polyethylene boxes were used to protect the DAQ board from the proton beam during the SEE test.} 
\label{fig:test_field} 
\end{figure}

 
The test setup configuration is shown in Figure~\ref{fig:sys_bd}. The analog signal fed into the ADC is generated by an RF signal generator SG384 from Stanford Research Systems, and it is a sine wave signal. The frequency of this sine wave signal is set as 39.0625 KHz to get 1024 sampling points for one wavelength of the signal with 40 MHz sampling frequency. The sine wave signal is then split by the 1:16 analog signal splitter, and fed into the 16 analog channels of the ADC under test. Two Agilent 34972A power supplies were used to provide separate power to the ADC during the test. These could be controlled and monitored through the GPIB interface. The ADC board and the DAQ board were connected by a 10-inch FMC extension cable (see Figure~\ref{fig:test_field}). The DAQ board was kept far away from the center of the radiation so it could be shielded separately.
\par 
The host computer was connected to the router with a long Ethernet cable, and was placed in the control room, which is free from radiation.
\end{multicols}

\begin{multicols}{2}

\section{Test Results}
\label{sec:results}
TID tests of two samples of ADS52J90 and one sample of AD9249 have been performed. For the test of the two ADS52J90 samples, different data output interfaces were used during the TID test. 
The dose rate was set at 10 kRad(Si)/hour, and each test was planned for 30 hours since the test target is 300 kRad(Si). 
The TID test of the AD9249 was stopped at 150 kRad(Si) after the device stopped working after 110 kRad(Si) of irradiation. 
For the two ADS52J90 samples, the test with the JESD204B interface completed at 300 kRad(Si) as planned but the test with the LVDS interface was extended to 700 kRad(Si).

\par
Figure~\ref{fig:ads52j90_tid_lvds} and Figure~\ref{fig:ads52j90_tid_jesd204b} show the current consumption plots of the three power rails of the ADS52J90 with the two different interfaces through the TID test. Current increases are less than 5\% after 300 kRad(Si) irradiation for both ADS52J90 samples, working with LVDS and JESD204B data interfaces respectively. The power rail DVDD is the core power of the JESD204B serializer and the power rail LVDD is used to power the LVDS driver of ADS52J90 \cite{4}, so the power consumption of these two power rails are different when the device is working with different data interfaces. 
A power recycle was performed at the 55 kRad(Si) point, and the performance of the ADC was unchanged.
\begin{center}
\centering
		\includegraphics[width={0.5\textwidth}]{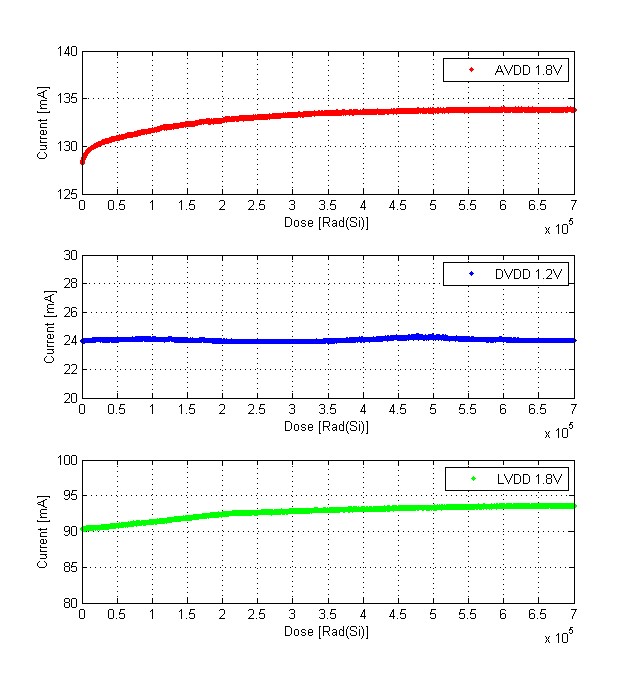}
		\figcaption{\label{fig:ads52j90_tid_lvds} Current consumption change of ADS52J90 working with LVDS interface.}
\end{center}

\begin{center}
\centering
		\includegraphics[width={0.5\textwidth}]{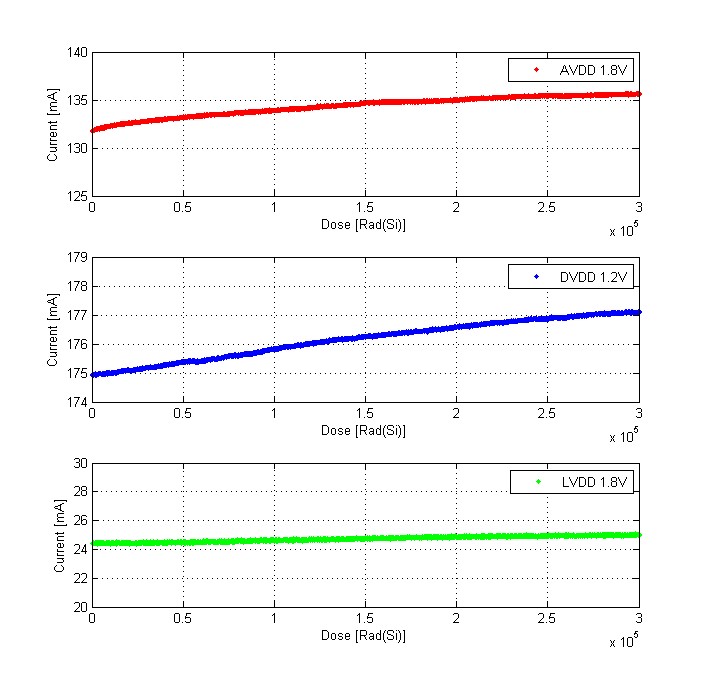}
		\figcaption{\label{fig:ads52j90_tid_jesd204b} Current consumption change of ADS52J90 working with JESD204B interface.}
\end{center}

Figure~\ref{fig:ad9249_tid} shows the current consumption plot of two power rails of the ADC AD9249. The recorded data shows that the analog power increased significantly after 50 kRad(Si) of irradiation. In addition, the device could no longer work after being irradiated for  about 110 kRad(Si). This result is similar to previous TID test results \cite{10} of ADCs from Analog Devices.\par

The SEE test modules were running throughout the TID test. No bit-flip errors or function failures were observed during the TID tests of both ADCs.

\begin{center}
		\includegraphics[width={0.5\textwidth}]{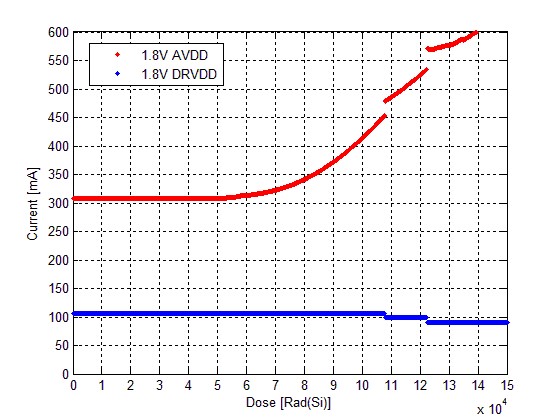}
		\figcaption{\label{fig:ad9249_tid}Current consumption change of AD9249 during the TID test.}
\end{center}
		
\end{multicols}
\begin{figure}
\centering 
\subfigure[~]{\includegraphics[width={0.4\textwidth}]{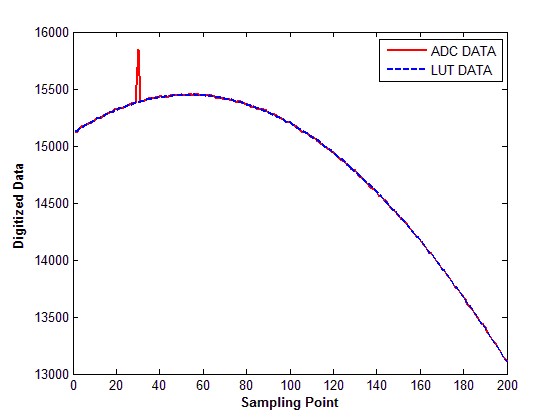}\label{fig:ads52j90_seu_sp} }
\subfigure[~]{\includegraphics[width={0.4\textwidth}]{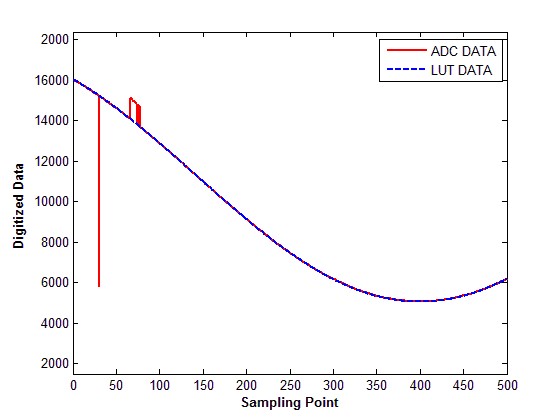} \label{fig:ads52j90_seu_mp} } 
\subfigure[~]{\includegraphics[width={0.4\textwidth}]{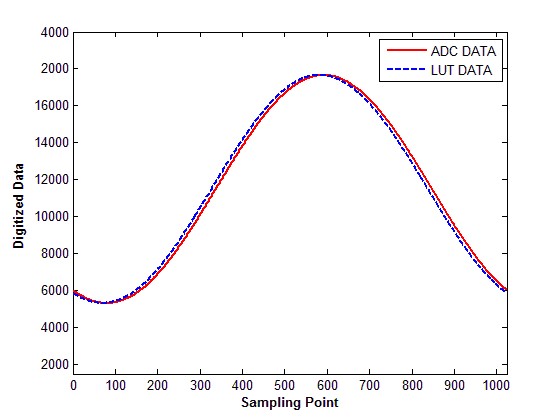}\label{fig:ads52j90_sefi_phs} }
\subfigure[~]{\includegraphics[width={0.4\textwidth}]{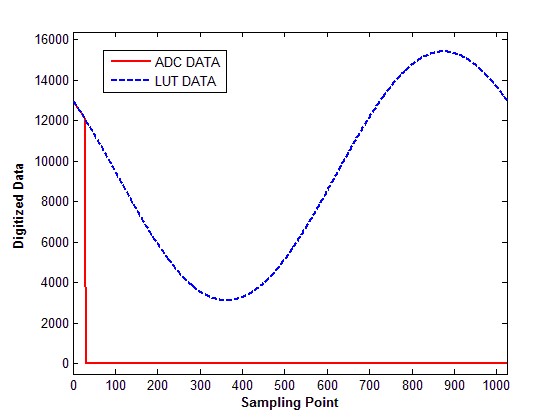} \label{fig:ads52j90_sefi_pd} } 
\caption{Typical SEE events of ADS52J90 during the SEE test: (a) Single-point SEU event; (b) Multi-point SEU event; (c) Phase change SEFI-A event; (d) Power down SEFI-A event.} 
%\label{fig:adc_board_design} 
\end{figure}
\begin{multicols}{2}
Since the TID test results of the AD9249 show it can survive only a relatively low total dose, only two samples of ADC ADS52J90 were tested during the SEE test at MGH. One sample operated with the LVDS data output interface and the other with the serial JESD204B interface.\par
The most common kind of SEE event observed during the SEE test is SEU. Figure~\ref{fig:ads52j90_seu_sp} shows a typical single point SEU event and Figure~\ref{fig:ads52j90_seu_mp} a multi-point SEU event. In both situations, the ADC can automatically be restored to its normal operational status quickly.\par
SEFI events did not occur as frequently as SEU events, and only type A SEFI events were observed during the SEE test of the ADC ADS52J90. The ADC cannot recover itself from the SEFI state, and needs a hardware reset through the dedicated pin of the ADC. Figure~\ref{fig:ads52j90_sefi_phs} shows a phase shift event and Figure~\ref{fig:ads52j90_sefi_pd} shows a power down event, both typical SEFI-A events.\par

Other types of SEFI-A events were observed during the SEE test of ADS52J90, including gain change and SPI configuration interface malfunction.\par 
After analysis of all the offline data of the SEE test, the cross sections of SEU and SEFI-A events were determined (see Table~\ref{tab:i}). The SEE cross sections of the ADS52J90 working with LVDS and JESD204B interfaces are very close, but the cross section of SEFI-A events is much higher for the ADC with the JESD204B interface. This is reasonable since the digital part of the serial transmission of the JESD204B link is much more complicated than the LVDS transmission.
\begin{center}
	\tabcaption{\label{tab:i} Measured cross sections of the SEE events of ADS52J90 in $cm^{-2}$ per device.}
	\footnotesize
	\begin{tabular}{ccc}
		\toprule SEE Type & LVDS Output & JESD204B Output\\
		\hline
		SEU  & $3.6 \times 10^{-10}$ & $3.1 \times 10^{-10}$\\
		SEFI-A  & $6.7 \times 10^{-12}$ & $8.4 \times 10^{-11}$\\
		SEFI-B  & - & -\\
		\bottomrule
	\end{tabular}%
\end{center}

\section{Conclusion}
\label{sec:conclusion}
In this paper, the design of an ADC radiation tolerance characterization system for the Upgrade of the ATLAS LAr Calorimeter is presented. Two COTS ADC test boards have been designed and tested using this system. Various interesting radiation tolerance features of these two ADCs during the TID test and SEE test have been observed. 
\par
According to the radiation tolerance criteria for COTS (single-lot) devices of the LAr electronics for operation at HL-LHC for a total luminosity of 3000 $fb^{-1}$, the estimated integrated ionization dose with a safety factor of 7.5 is 250 kRad(Si) and the high-energy ($E>20 MeV$) hadron fluence with a safety factor of 2 is $3.8 \times 10^{12}$ $h/cm^{2}$ \cite{12}. 
\par
If we only consider the TID requirements of the LAr electronics, the ADS52J90 working with either the LVDS or JESD204B interface is appropriate for the LTDB application. But a total of 64 barrel LTDB boards (giving a total of 1280 16-channel ADCs) will be installed into the ATLAS calorimeter, and the expected number of SEU and SEFI-A events in an LHC operation fill period (10 hours, the estimated integrated luminosity is 2 $fb^{-1}$) are 1167 and 22 respectively if the measured cross sections of SEE and SEAF-A for ADS52J90(LVDS) in Table~\ref{tab:i} are applied. These rates are relatively higher than the experiment's requirements, therefore neither of the ADCs presented in this paper are qualified for the LTDB application.
\par
The test results show that this characterization system works well and is flexible for multiple-channel high-speed ADCs with LVDS or JESD204B interfaces. It is applicable to other collider physics experiments where characterization for radiation tolerant ADCs is required.

%\section{References}

%References should be formatted as follows:

%Journal: Author(s), Journal name, volume number (Issue No.):
%page number (year of publication) (as shown in Ref.~\citep{lab1})

%Monographs: Author(s), Title, Edition (Location of publisher: Publisher, publication year), page no. (as shown in Ref.~\citep{lab2})

%Collection: Author(s), Text title, Editor (Location of publisher: Publisher, Publication year), page no. (as shown in Ref.~\citep{lab3})

\acknowledgments{We would like to thank Mr. E. Cascio at the MGH facility for the excellent service and help. His expert guidance was important in the execution of the SEE test. We also acknowledge the help from Mr. August Hoffmann for the mechanical assembly and shielding implementation of test boards during the TID and SEE test.}

\end{multicols}

\vspace{15mm}

\vspace{-1mm}
\centerline{\rule{80mm}{0.1pt}}
\vspace{2mm}

\begin{multicols}{2}

\end{multicols}

\clearpage
%\end{CJK*}
\end{document}